\documentclass[12pt]{article}
\usepackage{graphicx}

\newcommand\pubnumber{WSU--HEP--XXYY}
\newcommand\pubdate{\today}

\def\wayne{School of Physics \\
Nankai University, Tianjin, 300071, P.R. China}
\def\support{\footnote{Supported by the National Natural Science 
Foundation of China (NSFC) under Contracts No. 11005061 and No. 11475090.}}

\def\Title#1{\begin{center} {\Large #1 } \end{center}}
\def\Author#1{\begin{center}{ \sc #1} \end{center}}
\def\Address#1{\begin{center}{ \it #1} \end{center}}

\newcommand\pubblock{\rightline{\begin{tabular}{l} \pubnumber\\
         \pubdate  \end{tabular}}}
\newenvironment{Abstract}{\begin{quotation}  }{\end{quotation}}
\newenvironment{Presented}{\begin{quotation} \begin{center} 
             PRESENTED AT\end{center}\bigskip 
      \begin{center}\begin{large}}{\end{large}\end{center} \end{quotation}}
\def\Acknowledgements{\bigskip  \bigskip \begin{center} \begin{large}
             \bf ACKNOWLEDGEMENTS \end{large}\end{center}}

\def\beq{\begin{equation}}
\def\eeq#1{\label{#1}\end{equation}}
\def\eeqn{\end{equation}}

\def\beqa{\begin{eqnarray}}
\def\eeqa#1{\label{#1}\end{eqnarray}}
\def\eeqan{\end{eqnarray}}

\let\bar=\overbar

\def\Dslash{\not{\hbox{\kern-4pt $D$}}}
\def\dslash{\not{\hbox{\kern-2pt $\del$}}}

\def\msb{{\bar{\ssstyle M \kern -1pt S}}}

\begin{document}
\begin{titlepage}
\pubblock

\vfill
\Title{D rare/forbidden decays at BESIII}
\vfill
\Author{Minggang Zhao\support (For the BESIII Collaboration)}
\Address{\wayne}
\vfill
\begin{Abstract}
In this document we present the latest result on rare/forbidden decays for D mesons 
at the BESIII experiment. Based on 2.92 fb$^{−1}$ data taken at the center-of-mass 
energy 3.773 GeV with the BESIII detector, the flavor-changing neutral current process 
$D^0\to\gamma\gamma$ is searched using a double tag technique, while the decays of 
$D^+\to h^{\pm}e^+e^{\mp}$ ($h$ stands for $K$ or $\pi$) are studied based on a single 
tag method. The resulting upper limits are still above the Standard Model predictions.
\end{Abstract}
\vfill
\begin{Presented}
The 7th International Workshop on Charm Physics (CHARM 2015)\\
Detroit, MI, 18-22 May, 2015
\end{Presented}
\vfill
\end{titlepage}
\def\thefootnote{\fnsymbol{footnote}}
\setcounter{footnote}{0}
%

\section{Introduction}

One way to search for physics beyond the Standard Model is to search 
for decays that are forbidden or predicted to occur at a negligible level. 
Observing such decays would constitute evidence for new physics, and 
measuring their branching fractions would provide insight into how to 
modify our theoretical understanding.
For example, the absence of flavor-changing neutral currents (FCNCs) in 
kaon decays led to the prediction of the charm quark \cite{ref::cleo1}, 
and the observation of $B^0-\bar B^0$ mixing, a FCNC process, indicated 
a very large top-quark mass \cite{ref::cleo2}. Till now, the rare and 
forbidden charm decays have been less informative and less extensively 
studied.

FCNC processes in charm decays are highly suppressed by the 
Glashow-Iliopoulos-Maiani (GIM) mechanism \cite{ref::gim}, and can only 
occur via higher-order diagrams within SM , but the estimated branching 
fractions are $10^{-8}$ to $10^{-6}$ \cite{ref::fcnc}.
Such a small branching fractions are touching the sensitivity of current 
experiments. However, if additional new particles or mechanism exist, they 
could contribute additional amplitudes that would make these modes observable.
Thus, the hints of $D^+$ FCNC decays might provide indication of non-SM physics 
or of unexpectedly large rates, such as at $10^{-5}$ or $10^{-6}$ level \cite{ref::long-distance}, 
for long-distance SM processes $D^+\to\pi^+ V$, $V\to e^+e^-$, with a real or 
virtual vector meson $V$ (can be a $\rho$, $\omega$, or $\phi$).
The LNV decays $D^+\to K^-e^+e^+$ and $D^+\to\pi^-e^+e^+$ are strictly 
forbidden in the SM. They could be induced by a Majorana neutrino, but with 
a branching fraction only of order $10^{-23}$ \cite{ref::majorana}. So any 
observation at experimentally accessible levels would be clear evidence of new physics.
The searches for these decay modes have been carried out in several experiments 
(see table \ref{tab::historical}). 
In this document, we present latest results of searching for the FCNC 
decays of $D^0\to\gamma\gamma$, $D^+\to K^+e^+e^-$ and $D^+\to\pi^+e^+e^-$ 
together with the lepton-number violating (LNV) decays of $D^+\to K^-e^+e^+$ 
and $D^+\to\pi^-e^+e^+$ at the BESIII experiment. 

\begin{table*}[htbp]
\begin{center}
\caption{Comparisons of the upper limits ($10^{-6}$) on the branching fractions for $D^+\to h^{\pm}e^{\mp}e^{+}$ at a 90\% C.L..}\label{tab::historical}
\begin{tabular}{lccccccc}
\hline
Experiments & $D^+\to K^+e^+e^-$ & $D^+\to K^-e^+e^+$ & $D^+\to\pi^+e^+e^-$ & $D^+\to \pi^-e^+e^+$ \\
\hline
CLEO \cite{ref::cleo-1988}  &   -  &   -  & 2600 &   -  \\
MARK2 \cite{ref::mark-1990} & 4800 & 9100 & 2500 & 4800 \\
E687 \cite{ref::e687-1997}  & 200  & 120  & 110  & 110  \\
E791 \cite{ref::e791-1999}  & 200  & -    & 52   & 96   \\
CLEO \cite{ref::cleo-2010}  & 3.0  & 3.5  & 5.9  & 1.1  \\
Babar \cite{ref::babar-2011}& 1.0  & 0.9  & 1.1  & 1.9  \\
PDG \cite{ref::pdg2014}     & 1.0  & 0.9  & 1.1  & 1.1  \\
\hline
This work                  & 1.2  & 0.6  & 0.3  & 1.2 \\
\hline
\end{tabular}
\end{center}
\end{table*}

\section{Technique}

For the $e^+e^-$ annihilation experiment around the $\psi(3770)$ peak, 
the $D$ mesons are produced in pairs, i.e., if a $D$ meson is reconstructed 
in an event, which is called a {\it singly tagged $D$ event}, there must exist 
a $\bar D$ meson in the recoiling side. If the pair $D\bar D$is fully 
reconstructed in an event, the event is called a {\it doubly tagged $D\bar D$ event}. 
Traditionally, there are two methods to perform the searching for the rare/forbidden 
decays. One is based on singly tagged events which will provide large statics with 
high backgrounds, another is using doubly tagged events presenting extremly low
backgrounds while bad statics (see table \ref{tab::technique}). 
On which technique a searching will employ depends on both background contaminations 
and the statistics.

\begin{table*}[htbp]
\begin{center}
\caption{Two techniques on rare/forbidden searching.}\label{tab::technique}
\begin{tabular}{lccccccc}
\hline
Method & Statistics (charged/neutral) & Background & Sensitivity \\
\hline
Single Tag Method & $1.7\times10^7$/$2.1\times10^7$ & not good  & Bkg. vs Stat. \\
Double Tag Method & $1.6\times10^6$/$2.8\times10^6$ & clean     & Bkg. vs Stat. \\
\hline
\end{tabular}
\end{center}
\end{table*}

\section{$D^+\to h^{\pm}e^+e^{\mp}$}
\begin{figure*}[htbp]
\begin{center}
\includegraphics[height=9cm,width=14cm]{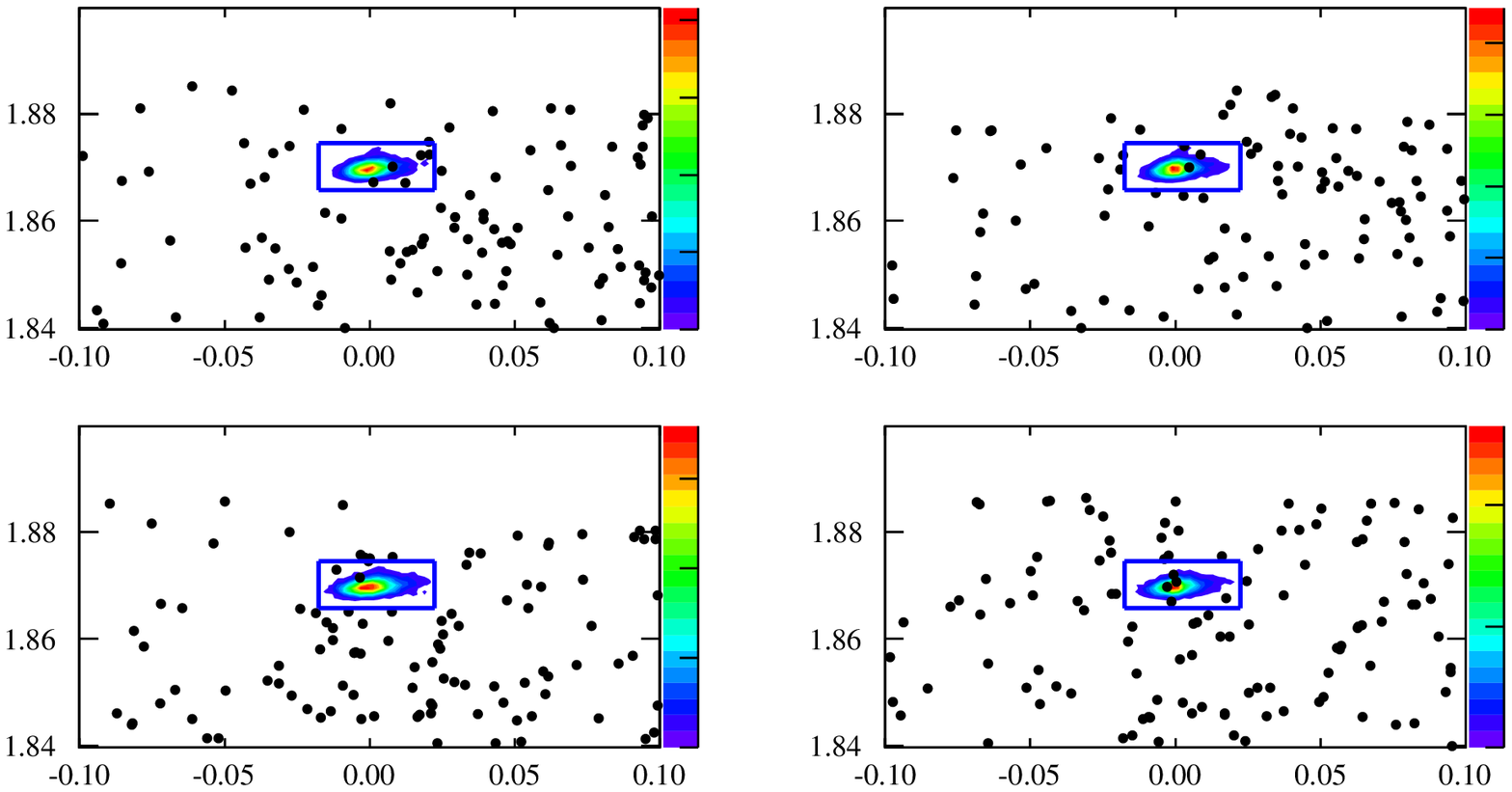}
\put(-220,-10){\bf \large $\Delta E$ [GeV]}
\put(-400,90){\rotatebox{90}{\bf \large $M_{\rm BC}$ [GeV/$c^2$]} }
\put(-325,235){\bf $D^+\to K^+e^+e^-$} 
\put(-330,225){\bf BESIII preliminary} 
\put(-125,235){\bf $D^+\to \pi^+e^+e^-$}
\put(-130,225){\bf BESIII preliminary}
\put(-325,107){\bf $D^+\to K^-e^+e^+$}
\put(-330,97){\bf BESIII preliminary}
\put(-125,107){\bf $D^+\to \pi^-e^+e^+$}
\put(-130,97){\bf BESIII preliminary}
\caption{The scatter plots of $M_{\rm BC}$ versus $\Delta E$ of the accepted events in data,
where the blue rectangle denotes the signal box.
The contour plot is determined by MC simulations. The scale of the MC is arbitrary.} \label{fig::box-data}
\end{center}
\end{figure*}

To searching the decays of $D^+\to h^{\pm}e^+e^{\mp}$, where $h$ means $K$/$\pi$, we 
check the accepted events in the signal boxes in the scatter plots of $M_{\rm BC}$ 
versus $\Delta E$ which are shown in figure \ref{fig::box-data}.
The signal box, which are kept blind before cuts optimizing and backgrounds studies
based on MC and sideband data, is deined with the mean value and the resolution of $M_{\rm BC}$ 
and $\Delta E$ determined from MC simulations, i.e.
$|\Delta E -\Delta E_{\rm mean}|<3\sigma_{\Delta E}$
and
$|M_{\rm BC}- M_{D^+}| <3\sigma_{M_{D^+}}$,
where $\Delta E_{\rm mean}$ is the mean value of the $\Delta E$ distribution,
$M_{D^+}$ is the $D^+$ nominal mass,
$\sigma_{\Delta E}$ and $\sigma_{M_{D^+}}$ are the corresponding resolutions.
Events falling into the signal box, shown as blue rectangle in figure \ref{fig::box-data},
are taken as candidate signal events.

Table \ref{tab::numbers} summarizes the numbers of events inside 
($N^{\rm data}_{\rm inside}$) and outside ($N^{\rm data}_{\rm outside}$) 
the signal boxes, the scale factors ($f_{\rm scale}$), detection efficiencies ($\epsilon$), 
systematic uncertainties ($\Delta_{\rm sys}$), calculated upper limits for 
the observed events ($\mathcal{s}_{90}$) and branching fractions ($\mathcal{B}$). 
The upper limits on the numbers of these decays is determined by utilizing a 
frequentist method \cite{ref::TROLKE} with unbounded profile likelihood treatment 
of systematic uncertainties, where the number of the observed events is assumed 
to follow a Poisson distribution, the number of background events and the efficiency 
are assumed to follow Gaussian distributions, and the systematic uncertainty is 
considered as the standard deviation of the efficiency.
Our results for $D^+\to\pi^+e^+e^-$ and $D^+\to K^-e^+e^+$ are significantly
improved than the previous restrictions and the other two limits are comparable
with the world best results.

\begin{table}[htbp]
\begin{center}
\caption{Summary of the numbers.}
\label{tab::numbers}
\begin{tabular}{ccccccccccc}
\hline
 & $N^{\rm data}_{\rm inside}$ & $N^{\rm data}_{\rm outside}$ & $f_{\rm scale}$ 
 & $\epsilon$ [\%] & $\Delta_{\rm sys}$ [\%] & $\mathcal{s}_{90}$ & $\mathcal{B}$ [$\times10^{-6}$] \\
\hline
$D^+\to K^+e^+e^-$   & 5 & 69 & $0.08\pm0.01$ & 22.53 & 5.4 & 19.4 & $<1.2$ \\
$D^+\to K^-e^+e^+$   & 3 & 55 & $0.08\pm0.01$ & 24.08 & 6.1 & 10.2 & $<0.6$ \\
$D^+\to \pi^+e^+e^-$ & 3 & 65 & $0.09\pm0.02$ & 25.72 & 5.9 & 4.2  & $<0.3$ \\
$D^+\to \pi^-e^+e^+$ & 5 & 68 & $0.06\pm0.02$ & 28.08 & 6.8 & 20.5 & $<1.2$ \\
\hline
\end{tabular}
\end{center}
\end{table}

\section{$D^0\to\gamma\gamma$\cite{ref::BESIII-2gamma}}
To suppress the backgrounds from QED continuum processes, potential 
$\psi(3770)\to {\rm non}-D\bar D$ decays, as well as $D^+D^−$ decays, 
we perform a double tag technique in the analysis. 
In this work, singly tagged events are selected as the first step, then 
$\gamma\gamma$ final states will be investigated in the system recoiling side.

Single tag candidates are selected by reconstructing a $\bar D^0$ in one of the following 
five hadronic final states: $\bar D^0\to K^+\pi^−$, $K^+\pi^−\pi^0$, $K^+\pi^−\pi^+\pi^−$,
$K^+\pi^−\pi^+\pi^−\pi^0$, and $K^+\pi^−\pi^0\pi^0$, constituting approximately 37\% of 
all $D^0$ decays \cite{ref::pdg2012}.
The absolute branching fraction for the signal mode is determined as,
\begin{equation}
\mathcal{B}=\frac{N_{\rm tag, \gamma\gamma}}{\Sigma_{i}N^{i}_{\rm tag}\cdot(\epsilon^{i}_{\rm tag, \gamma\gamma}/\epsilon^{i}_{\rm tag})},
\end{equation}
where $i$ runs over each of the five tag modes, $N_{\rm tag}$ and $\epsilon_{\rm tag}$ are 
the single tag yield and reconstruction efficiency, and $N_{\rm tag, \gamma\gamma}$ and 
$\epsilon^{i}_{\rm tag, \gamma\gamma}$ are the yield and efficiency for the double tag 
combination of a hadronic tag and a $D^0\to\gamma\gamma$ decay.

\begin{figure}[htb]
\centering
\includegraphics[height=9cm,width=12cm]{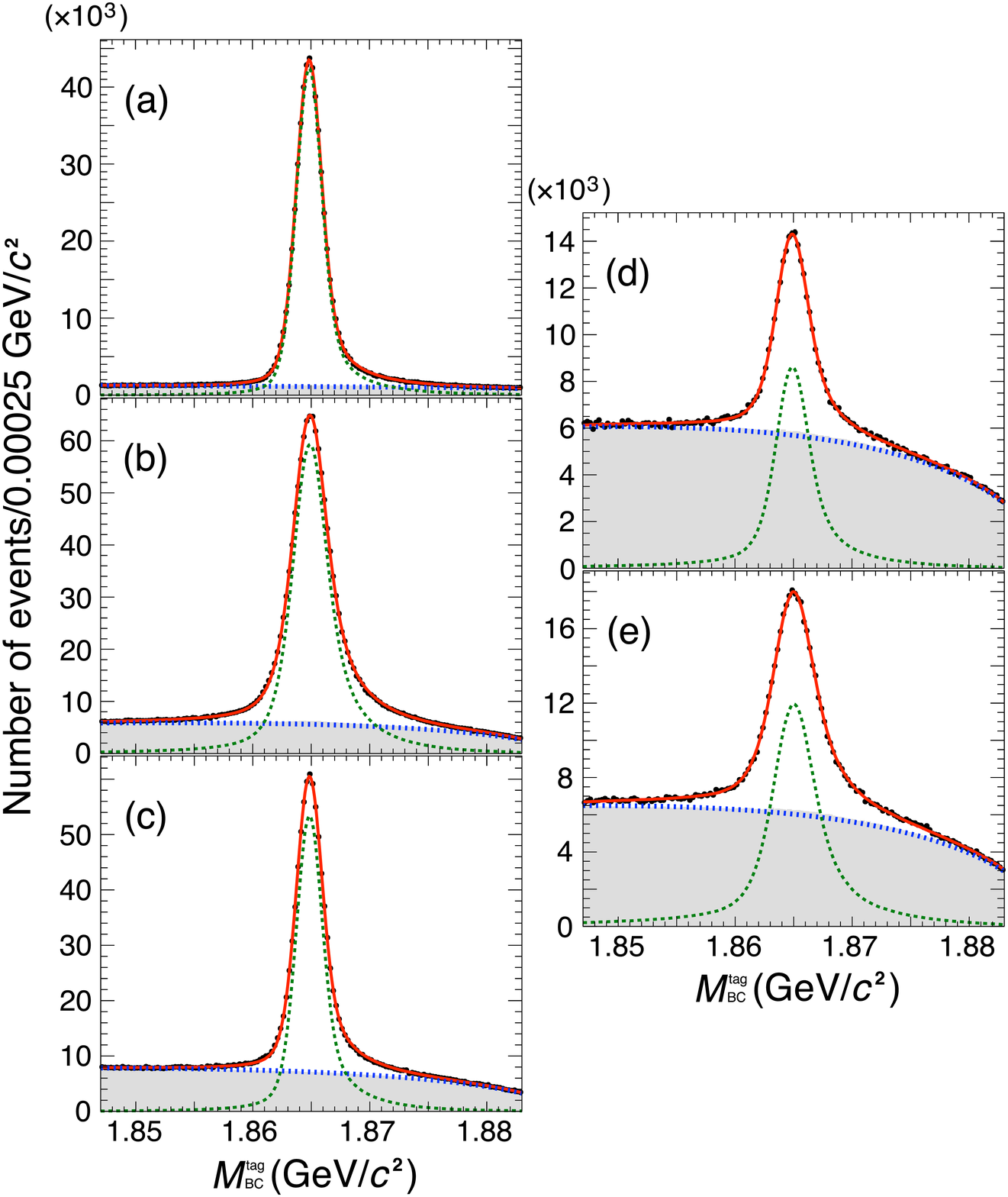}
\caption{Fits (solid line) to the $M^{\rm tag}_{\rm BC}$ distributions 
in data (points) for the five $D^0$ tag modes: (a) $K^+\pi^−$, (b) $K^+\pi^−\pi^0$, 
(c) $K^+\pi^−\pi^+\pi^−$, (d) $K^+\pi^−\pi^+\pi^−\pi^0$, and (e) $K^+\pi^−\pi^0\pi^0$. 
The gray shaded histograms are arbitrarily scaled generic MC backgrounds.}
\label{fig::tags}
\end{figure}

We extract the single tag yield for each tag mode and the combined yields 
of all five modes from fits to $M^{\rm tag}_{\rm BC}$ distributions. 
The signal shape is derived from the MC simulation which includes the effects of 
beam-energy smearing, initial-state radiation, the $\psi(3770)$ line shape, and detector 
resolution. We then convolute the line shape with a Gaussian to compensate for a difference 
in resolution between data and our MC simulation. Mean and width of the convoluted Gaussian, 
along with the overall normalization, are left free in our nominal fitting procedure. 
The background is described by an ARGUS function \cite{ref::argus}, which models combinatorial 
contributions. In the fit, all parameters of the background function are left free, except 
its endpoint which is fixed at 1.8865 GeV$/c^2$.
Figure \ref{fig::tags} shows the fits to our tag-candidate samples. 

\begin{figure}[htb]
\centering
\includegraphics[height=9cm,width=7cm]{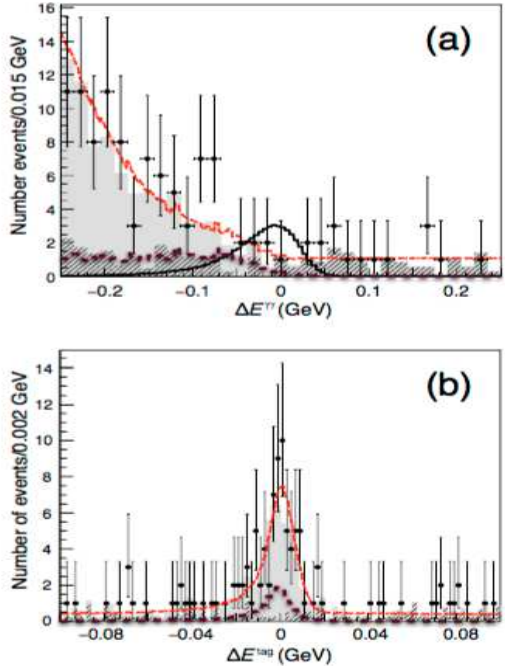}
\caption{Fit to the DT sample in data (points), projected onto $\Delta E^{\gamma\gamma}$ (a)
and $\Delta E^{\rm tag}$ (b). The dashed lines show the overall fits, while the dotted 
histograms represent the estimated background contribution from $D^0\to\pi^0\pi^0$. 
The solid line superimposed on the ∆Eγγ projection indicates the expected signal for 
$\mathcal{B}(D^0\to\gamma\gamma)=10\times10^{−6}$. Also overlaid are the overall MC-estimated 
backgrounds (gray shaded histograms) and the background component from non-$D\bar D$ processes 
(diagonally hatched histograms).}
\label{fig::sig-tag}
\end{figure}

Although we can suppress most of the background with the double tag method, 
there remain residual contributions from continuum processes, primarily 
doubly-radiative Bhabha events for $Kπ$ tags and $e^+e^−\to q\bar q$ for 
other modes. In order to correctly estimate their sizes, we take a data-driven 
approach by performing an unbinned maximum likelihood fit to the two-dimensional 
distribution of $\Delta E^{\gamma\gamma}$ versus $\Delta E^{\rm tag}$, as shown 
in figure \ref{fig::sig-tag}.
We use $\Delta E^{\gamma\gamma}$ distributions rather than $M^{\gamma\gamma}_{\rm BC}$
distributions as the background from non-$D\bar D$ decays is more easily addressed 
in the fit. Also, the background from $D^0\to\pi^0\pi^0$ peaks in 
$M^{\gamma\gamma}_{\rm BC}$ at the same place as the signal does, whereas it is 
shifted in $\Delta E^{\gamma\gamma}$. 
The fitting ranges are $|\Delta E^{\gamma\gamma}| < 0.25$ GeV and 
$|\Delta E^{\rm tag}| < 0.1$ GeV. These wide ranges are chosen to have adequate 
statistics of the continuum backgrounds in our fit. 
For the signal and the $D\to\pi^0\pi^0$ background, we extract probability 
density functions (PDFs) from MC simulations. For the background from continuum 
processes, we include a flat component in two dimensions, allowing the 
normalization to float. The contribution from $D^+D^−$ decays is completely 
negligible. We model the background from other $D^0\bar D^0$ decays with a pair 
of functions. In the $\Delta E^{\rm tag}$ dimension we use a Crystal Ball 
function (CB) \cite{ref::Crystal-Ball} plus a Gaussian, and in the 
$\Delta E^{\gamma\gamma}$ dimension, we use a second-order exponential polynomial
function.
In our nominal fitting procedure, we fix the following parameters based on MC: 
the power-law tail parameters of the CB, the coefficients of the polynomial, and 
the mean and the width of the Gaussian function. 
The normalization for the background from all other $D^0\bar D^0$ decays is left 
free in the fit, as are the mean and width of the CB and the ratio of the areas 
of the CB and Gaussian functions. 
Figure \ref{fig::sig-tag} shows projections of the fit to the DT data sample
onto $\Delta E^{\gamma\gamma}$ (top) and $\Delta E^{\rm tag}$ (bottom).

The fit yields $N_{\rm tag, \gamma\gamma} = (−1.0^{+3.7}_{-2.3})$, demonstrating that 
there is no signal for $D^0\to\gamma\gamma$ in our data. This corresponds to 
$\mathcal{B}(D^0\to\gamma\gamma) = 3.8× 10^{−6}$ including the systematic uncertainties.
If the systematic uncertainty were ignored in setting this limit it would be reduced
by $0.1\times10^{−6}$. 

\section{Summary}

In summary, by analyzing 2.92 fb$^{-1}$ data collected at $\sqrt{s}=3.773$ GeV
with the BESIII detector at the BEPCII collider, we search for the
FCNC decays $D^0\to\gamma\gamma$, $D^+\to h^+e^+e^-$ and the LNV decays $D^+\to h^-e^+e^+$.
No signal excess is observed. As a result, we set the upper limits on the branching fractions
for these decays at a 90\% CL.
The results for $D^+\to\pi^+e^+e^-$ and $D^+\to K^-e^+e^+$ are significantly
improved than the previous restrictions, while those for $D^+\to\pi^-e^+e^+$ and $D^+\to K^+e^+e^-$ are comparable
with the world best results.
The result for $D^0\to\gamma\gamma$ is consistent with the upper limit previously set 
by the BABAR Collaboration \cite{ref::babar-2gamma}. 
Our result is the first experimental study of this decay using data at open-charm threshold, 
where the backgrounds from non-$D\bar D$ decays can be effectively suppressed.
The resulting upper limits are still above the Standard Model predictions, no hints for 
New Physics have been found yet.

\Acknowledgements
I would like to thank the committee of CHARM 2015 for the invitation and their excellent 
organizing. 


\end{document}